\documentclass{article}
\usepackage[dvips]{graphicx}
\def\sst{\scriptscriptstyle}

\def\ra{\rightarrow}

\def\al{\alpha}

\def\be{\begin{equation}}
\def\ee{\end{equation}}
\def\bea{\begin{eqnarray}}
\def\eea{\end{eqnarray}}

\input epsf.tex

\title{
\rightline{IFT-UAM/CSIC-99-29}
\rightline{hep-th/9907175}
\vspace{1cm}
T-duality and The Gravitational Description Of Gauge Theories\footnote{To appear in the Proceedings of the Schladming Winter School 1999.}}
\author{Cesar Gomez$^{\clubsuit\diamondsuit}$\footnote{E-mail:iffgomez@roca.csic.es} and Pedro Silva$^{\spadesuit \flat}$\footnote{E-mail:psilva@delta.ft.uam.es}\\
\\
\\
{\it $^{\clubsuit}$ Instituto de Fisica Teorica, C-XVI, Universidad Autonoma de Madrid}\\
{\it E-28049-Madrid, Spain\footnote{Unidad de Investigacion asociada al centrode Fisica Miguel Catalan (C.S.I.C.)}}\\
{\it $^{\diamondsuit}$I.M.A.F.F., C.S.I.C., Calle de Serrano 113}\\
{\it E-28006-Madrid, Spain}\\
{\it $^{\spadesuit}$Physics department, University of Newcastle Upon Tyne}\\
{\it NE1 7RU, UK}\\
{\it $^{\flat}$Depto de Fisica Teorica, C-XI, Universidad Autonoma de Madrid,}\\
{\it E-28049-Madrid, Spain}}

\begin{document}
\def\ft#1#2{{\textstyle{{\scriptstyle #1}\over {\scriptstyle #2}}}}
\def\sst{\scriptscriptstyle}

\maketitle
\begin{abstract}{This is a review of some basic features on the relation between supergravity and pure gauge theories with special emphasis on the relation between T-duality and supersymmetry. Some new results concerning the interplay between T-duality and near horizon geometries are presented.} 
\end{abstract}

\pagebreak
\newpage


\section{Introduction}

\begin{figure}
\begin{center}
\includegraphics[angle=-90,scale=0.4]{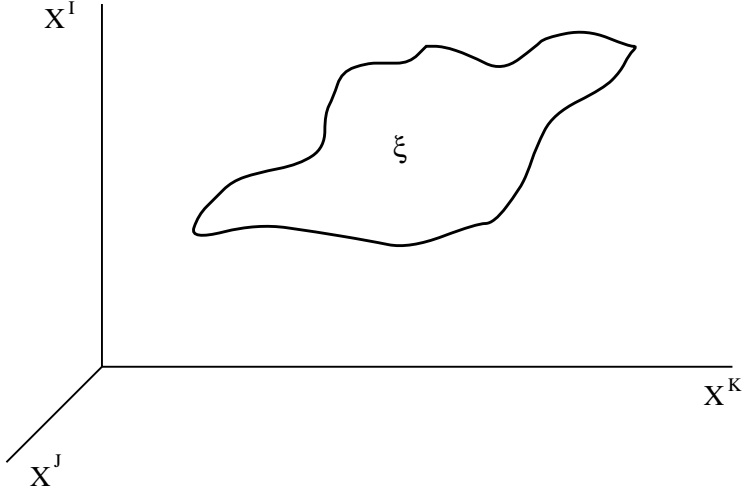}
\caption{Space-Time}
\label{fig:f1}
\end{center}
\end{figure}

String theory \cite{gsw,pol1,kis1} is defined by the two dimensional non linear sigma-model
\be
S = {1\over 4\pi\alpha'}\int\!\!d^2\xi\;\bigg\{
\left[\sqrt{-g}g^{ij}G_{\mu\nu} + \varepsilon^{ij}B_{\mu\nu}\right]\partial_iX^\mu\partial_jX^\nu + \alpha' \Phi\sqrt{-\gamma} R^{(2)}\bigg\},\ 
\label{sig}
\ee
provided the following space-time interpretations (fig. \ref{fig:f1}): 
\begin{enumerate}
\item $X$ are the space-time coordinates where the string is embed.
\item $G,\Phi,B$ are the external background fields called the metric, the dilaton and the torsion, respectively.
\item $g$ is the world sheet metric.
\item $\xi$ are the world sheet coordinates.
\end{enumerate}

The governing principle of string theory is the world sheet Weyl invariance of (\ref{sig}), understood as a two dimensional field theory. Interpreting $G,B$ and $\Phi$ as coupling constraints  the requirement of Weyl invariance becomes equivalent to the following equations:
\bea
\beta^G_{\al\beta}&=&R_{\al\beta}-\frac{1}{4}H^2_{\al\beta}+2\nabla_\al\nabla_\beta\Phi, \nonumber \\
\beta^B_{\al\beta}&=&\frac{1}{2}\nabla^\gamma H_{\gamma\al\beta}-\nabla^\gamma\Phi H_{\gamma\al\beta}H, \nonumber \\
\beta^\Phi&=&\frac{D}{6}+\frac{\al'}{2}[-R+\frac{H^2}{12}+4(\nabla\Phi)^2-4\nabla^2\Phi],
\label{beta}
\eea
where the $\beta$'s in (\ref{beta}) are the different beta functions for the Lagrangian of (\ref{sig}) and $R$,$H$ are the strength fields of $G$,$B$.

One of the most interesting aspects of string theory can be already obtained by direct inspection of the first equation of (\ref{beta}) which present a strong similarity with general relativity Einstein's equations. More precisely we can define associated with the string theory (\ref{sig}) an effective Lagrangian on the background fields ${\cal L}(G,B,\Phi)$ such that the corresponding equations of motions
\begin{figure}
\begin{center}
\includegraphics[angle=-90,scale=0.4]{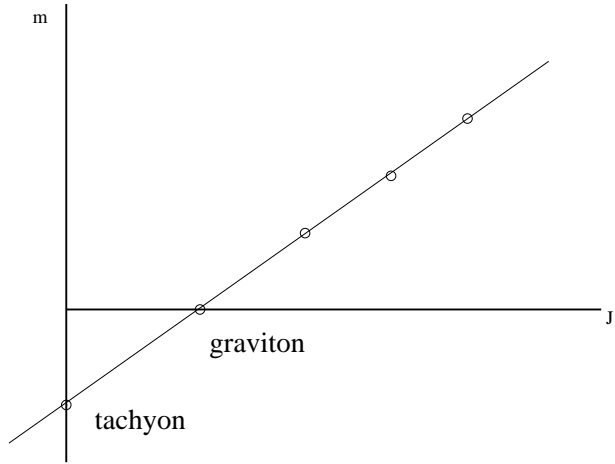}
\caption{Regge Trajectory}
\label{fig:f2}
\end{center}
\end{figure}
\be
{\delta{\cal L} \over \delta\Psi_i}=0\;\;\;,\;\;\;\Psi_i=G,B,\Phi,
\ee
coincide precisely with the beta function equation (\ref{beta}). This effective Lagrangian is given by:
\be
I={1\over 2\kappa^2}\int dx^{26}\sqrt{-G}e^{2\Phi}\Big[R+4\nabla_{\mu}\phi\nabla^{\mu}\Phi
-\frac{1}{12}H^2\Big]
\label{lagb}
\ee
Note that this lagrangian is not defined in the canonical form of Einstein general relativity. To make contact with the more familiar Einstein formalism we perform a rescaling on the metric absorbing the dilaton into the new metric. The resulting metric is known as the Einstein metric or the canonical metric, while the original metric is called string metric. The precise transformation is given by:
\be
G^e_{\al\beta}=e^{-\Phi/2}G^s_{\al\beta}.
\ee
   
The previous discussion correspond to the closed bosonic string. The physical spectrum of this string contains a tachyonic mode with mass square ${-2 \over  \al'}$, a massless state of spin two containing the graviton, antisymmetric tensor and the trace part that is identify with the dilaton. On the top of this, a tower of massive states with masses proportional to ${-1 \over  \al'}$ (fig.\ref{fig:f2}).

Graviton and tachyon scattering amplitudes for this string theory  are defined in terms of vertex operators:
\bea
V_{tachyon}&=& e^{ikx},   \nonumber \\
V_{graviton}&=&\eta_{\mu\nu}\partial X^\mu \partial X^\nu e^{ikx},
\eea

At tree level these amplitudes are consistently defined for $D\leq 26$, for $D$ the space-time dimension. At one loop level the requirement of unitarity implies that $D$ should be equal to the critical dimension $D=26$. 

\begin{figure}
\begin{center}
\includegraphics[angle=-90,scale=0.4]{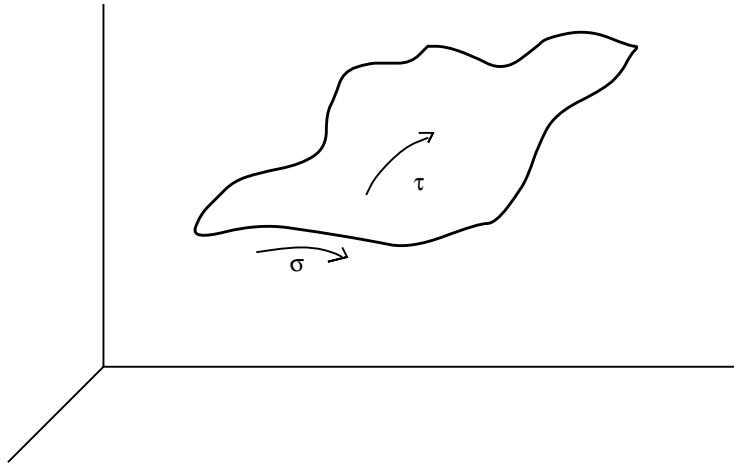}
\caption{World-sheet parametrisation}
\label{fig3}
\end{center}
\end{figure}

A slightly different type of strings are the open string. In this case the Regge trajectory contains a massless spin one  state that we can try to identify with some sort of gauge boson. The definition of open strings requires to specify precise boundary conditions at the end points of the string. If we want to preserved target space-time Lorenz invariance we should choose Newman boundary conditions.
\be
\partial_\sigma X=0\mid_{\hbox{{\small end points}}}
\ee
with the parameterization of the world sheet as indicated on (fig.\ref{fig3}).

\begin{figure}
\begin{center}
\includegraphics[angle=-90,scale=0.4]{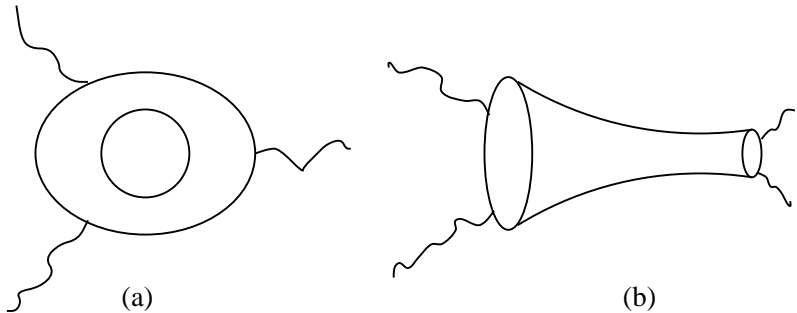}
\caption{One loop amplitudes}
\label{fig4}
\end{center}
\end{figure}

In addition the above type of strings allow us to decorate its ends with additional information, the so called Chan-Paton factors, that we can heuristically imagine as pairs of ``quark-antiquark'' transforming in the fundamental representation of some gauge group $G$. This makes for $G=U(N)$ that the open string states will transform in the adjoin representation as it should be for a gauge boson. In case we consider orthogonal groups $G=SO(N)$ the requirement for the gauge boson to transform in the adjoint representation implies to introduce an orientation to the open string. Gauge boson associated with exceptional algebras can not be included on this way. This was one of the main motivations for the discovery of the ``heterotic string'', that surprisely enough are closed strings that contains in its spectrum massless gauge bosons for the group $E_8\otimes E_8$.
Gauge boson amplitudes for the open bosonic string can be easily computed using the following vertex operators:
\be
V=\xi\partial Xe^{ikx}.
\ee
As in the case of the closed bosonic string, the open strings contains a tower on massive states, with masses of order $1/\al'$, and  a tachyon with negative square mass. In figure (\ref{fig4}) we have depicted some open string one loop amplitudes. Examples like (a) are planar, i.e. they can be draw in a plane, while examples as (b) can not, and are called non-planar.

One of the deepest aspects of open string theory can be already discussed from direct inspection of diagram (b). Namely, from the standard scattering theory point of view, what we are seeing is a scattering of open string with closed string states contributing to the internal channel. In other words closed strings appears naturally as interaction products of open strings. This simple and basic fact of string theory should immediately ring a bell of any quantum field theorist. In fact we can always work out string theory in the infinite tension limit ($\al' \ra 0$) where we decouple all the tower of massive states. From the open string point of view the result should be a pure gauge theory, while from the closed string point of view should land in pure gravity. We may wonder then, if there is any residual effect of the string open-closed relation that survives at the limit  $\al' \ra 0$?. This is a basic question that will allow us to enter into the very recent important developments connecting Yang Mill theories and gravity but before that we need to discuss another aspect of the open-closed interplay namely the D-branes and T-duality.


\section{D-Branes and T-Duality}


Let us star considering a closed string in a space time of the type $\Re^d\otimes S^1_R$, with $R$ the radius of the $S^1$. The extended nature of the closed string allowed us to define a new quantum number, namely the ``winding number'' of the closed string around the circle. Let us called it $m$. Now consider the mass formula for the closed string states after compactification on the circle $S^1$ (i.e. in our example masses from the point of view of the observers in $\Re^d$):
\be
M^2 = -p^\mu p_\mu = {2\over\al'}(\alpha_0^{25})^2+{4\over\al'}
({ N}-1)
\ee
and
\be 
\alpha^{25}_0 = \left({n\over R}+{mR\over\al}\right)
\sqrt{\al'\over 2},
\ee 
where $\mu$ runs only over the non-compact dimensions, $N$ is the
total level of the left-moving excitations.

This mass formula posses a very interesting and amazing symmetry defined by:
\bea
n\leftrightarrow m \;\;\;,\;\;\; R\leftrightarrow {\al' \over R}, 
\label{t1}
\eea
This symmetry is known as T-duality \cite{jap1}.

As it should be clear from the effective Lagrangian (\ref{lagb}) if we require not only invariance of the spectrum but also invariance for the amplitude we should change the dilaton field as 
\be
{\tilde \Phi}=  \Phi-ln({R \over \al'^{1/2}}).
\ee
The above example of T-duality can be generalized to generic backgrounds with a Killing vector. The T-duality transformation with respect to this isometry is given by the so called Buscher transformations \cite{bus1}:
\bea
{\tilde G}_{kk}&=&{1\over G_{kk}}, \nonumber\\
         {\tilde G}_{k\alpha}&=&{B_{k\alpha} \over G_{kk}},
\qquad
{\tilde B}_{k\alpha}={G_{k\alpha} \over G_{kk}}, \nonumber\\
          {\tilde G}_{\alpha\beta} &=& G_{\alpha\beta} -
{G_{k\alpha}G_{k\beta} - B_{k\alpha} B_{k\beta}\over G_{kk}}, \nonumber\\
        {\tilde
B}_{\alpha\beta}&=&B_{\alpha\beta}-{G_{k\alpha}B_{k\beta}
         -G_{k\beta}B_{k\alpha}\over G_{kk}},
\eea
where the letter $k$ stands for the direction of the isometry. Obviously the above transformations are generalisable to more than one isometry. Coming back to our initial example, we note that for the bosonic string $R\ra 0$ and $R\ra \infty$ are in all aspects equivalents provided we interchange the winding by the momenta. In practice what happen is that for instance in the limit $R=0$ an effective ``extra dimension'' appears due to the generation of massless winding modes.

For a while nobody ask, concerning this strange T-duality symmetry for closed strings, the most natural question, namely what is the interplay between T-duality and the already mentioned closed-open string relation?. 

Intuitively the problem we face is quite clear. In fact for the open string there is no winding number, so once we go to the limit $R\ra 0$ we most expect to end up with a $d$-dimensional theory (as usual in a quantum field theory), however open strings in interactions will produce closed strings and as we have seen in this case we get a new effective dimension, on the above limit. So what is really happening?
\begin{figure}
\begin{center}
\includegraphics[angle=-90,scale=0.4]{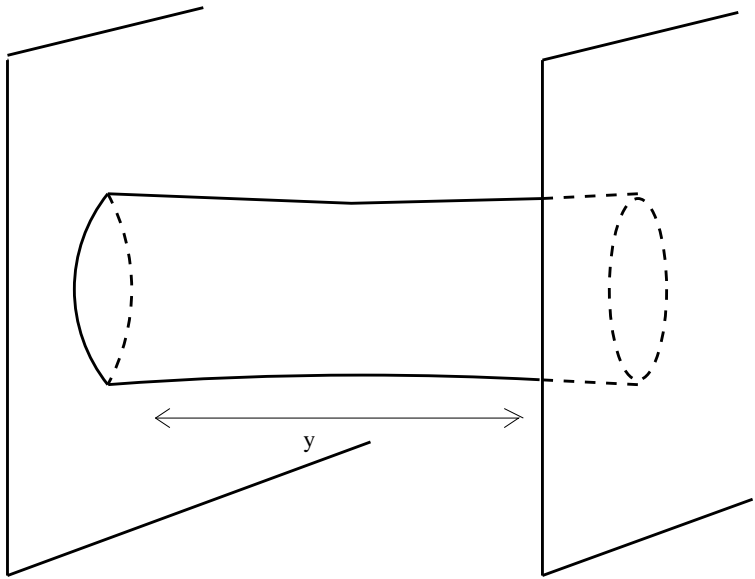}
\caption{D-brane interaction}
\label{fig5}
\end{center}
\end{figure}

The first answer to this puzzle is D-branes \cite{pol2}. As we will see in a moment open strings live in a $d+1$-dimensional space time, but with the end points attached to a $d$-dimensional region that we identify with the D-brane, more over the closed string in the $d+1$-dimensional space time will induce a gravitational life for the D-branes that will appears as a real source of gravity. Let us see all this in more detail. As we mentioned before open strings are characterized by the boundary conditions at the ends points of the string. The T-duality transformation of (\ref{t1}) is geometrically understand as mirror symmetry on one sector of the string mode expansion, this can be better seen by inspectioning the closed string expansion that solve the field equation,
\begin{eqnarray}
X^\mu&=&x^\mu  - i\sqrt{{\alpha'\over 2}} (\alpha^\mu_0 + \bar{ \alpha}^\mu_0 ) \tau
+ \sqrt{{\alpha'\over 2}} (\alpha^\mu_0 - \bar{\alpha}^\mu_0 )\sigma \nonumber \\
&+&i\sqrt{{\alpha'\over 2}} \sum_{m\ne 0} 
\left({\alpha^\mu_m\over m}z^{-m}+ {\bar{\alpha}^\mu_m\over m}\bar{z}^{-m}\right), 
\end{eqnarray}
where $z=e^{\tau-i\sigma}$ and $\bar{z}= e^{\tau+i\sigma}$. This expansion can be rewritten in terms of homomorphic ad anti-homomorphic functions as,
\be
X^\mu=X^\mu(z)+\bar{X}^\mu(\bar{z}),
\ee
then the T-duality transformation translates into a parity transformation for the anti-holomorphic function on the above expansion,
\begin{figure}
\begin{center}
\includegraphics[angle=-90,scale=0.4]{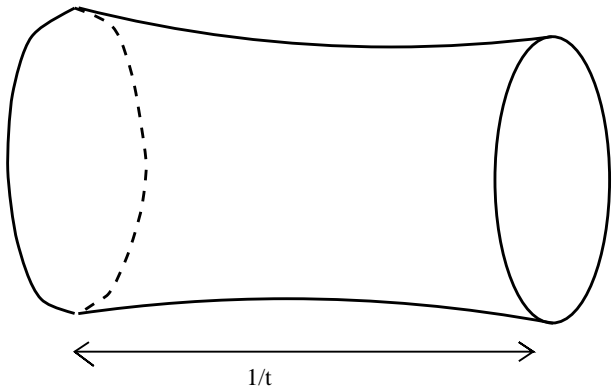}
\caption{Modular parameter}
\label{fig6}
\end{center}
\end{figure}
\be
{\tilde X^{25}}(z,\bar{z})=X^{25}(z)-\bar{X}^{25}(\bar{z}).
\ee
for the open string solution we have,
\bea
X^\mu&=&x^\mu-i\alpha' p^\mu \ln z\bar{z}+i\sqrt{{\alpha'\over 2}} \sum_{m\ne 0} {\alpha^\mu_m\over m}
(z^{-m}+\bar{z}^{-m}).
\eea
Therefore using the same transformation as in the closed string sector we get the expansion,
\bea
{\tilde X^{25}}&=&x^{25}-i\alpha' p^{25} \ln {z  \over \bar{z}}+i\sqrt{{\alpha'\over 2}} \sum_{m\ne 0} {\alpha^{25}_m\over m}
(z^{-m}-\bar{z}^{-m}).
\eea
Therefore the boundary condition $\partial_\sigma X^{25}=0$  is transformed into $\partial_\tau  X^{25}=0$, that means Dirichlet conditions at the ends of the open string. We can see that T-duality exchange Newman boundary conditions into Dirichlet boundary conditions. This implied that the ends of the open string are constraint to move on a given hypersurface, that we called D(irichlet)-branes.

At this point we can come back to our previous discussion concerning the filed theory limit $\al'\ra 0$. In fact if in this limit closed strings are still surviving them we should expect that T-duality in the field theory limit $\al'\ra 0$ of an open string will present somehow the phenomena of quantum generation of extra dimension. We will come back to this issue in chapter (8).

As a warming up exercise in D-brane dynamics let us consider the D-brane amplitude represented in (fig.\ref{fig5}). 
\begin{figure}
\begin{center}
\includegraphics[angle=-90,scale=0.4]{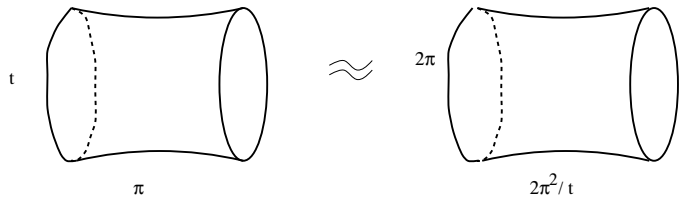}
\caption{Conformal mapping}
\label{fig7}
\end{center}
\end{figure}

Up to numerical actors that will not be relevant for us at this point the amplitude is given by:
\be
A(y)=(\alpha')^{-d/2}\int{dt \over t}t^{-d/2}\eta(it)^{(2-D)}e^{y^2t/\alpha'},
\label{eq:14}
\ee
where $\eta$ is the Dedekind function, $y$ the space-time distance between the Dp-brane, $d=p+1$ and $t$ the modular parameter of the cylinder (see fig.\ref{fig6}).

We can adopt two different points of view to interpret the amplitude in (fig.\ref{fig5}). From the open string point of view we have the open string stretched between the Dp-brane and a ``time'' evolution along the loop with the value of the time equal to $t$. From the closed string point of view we have the emission-absorption of closed string states with the ``time'' of the process given by $1/t$. Both pictures are related by the conformal mapping described in (fig.\ref{fig7}). We will be first interested in computing (\ref{eq:14}) in the limit $t \rightarrow 0$ that is the regime (see fig.\ref{fig6}) dominated by the contribution of light closed string states.

Using the transformation properties of the Dedeking function:
\be
\eta(-{1 \over it})=t^{-d/2}\eta(it)
\ee
and the expansion of $\eta$ for $-1/it \rightarrow \infty$ we get:
\be
A(y)=\left[(\alpha')^{-d/2}\int{dt \over t}t^{-d/2}t^{(2-D)/2}e^{y^2t/\alpha'}\right](D-2),
\ee
where $D$ is the dimension of the target space time and where  we have avoided the tachyon contribution in the expansion of the expansion of $\eta$-function \footnote{Recalled that:
\be
\eta(is)|_{is\rightarrow \infty}=(e^{1/24}(1+ e^{-25} + ...
\ee
with the factor $1$ representing the tachyon contribution.}

In critical dimension $D=26$ we get from the well known expression
\be
A(y)\approx \Gamma\left ({23-p \over 2}\right )|y|^{p-23}(\alpha')^{11-p}.
\label{eq:17}
\ee

\begin{figure}
\begin{center}
\includegraphics[angle=-90,scale=0.4]{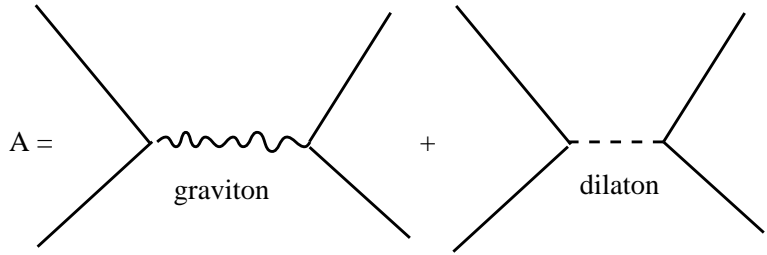}
\caption{Feynman diagrams}
\label{fig8}
\end{center}
\end{figure}

We can compare equation (\ref{eq:17})  with the effective Lagrangian computation. In fact (\ref{eq:17}) is a long distance contribution where we have keep only massless dilatons and gravitons, therefore we should compare (\ref{eq:17}) with the tree level Feynman diagrams for the effective Lagrangian \label{eq:4} in the Einstein frame and a D-brane coupled i.e. the Feynman diagram in (fig.\ref{fig8}). The coupling in the vertices in (fig.\ref{fig8}) will depend on the gravitational constant $\kappa$ and on the Dp-brane tension $\tau_p$ that we want to discover by identifying the amplitude in (fig.\ref{fig8}) and the amplitude in (fig.\ref{fig4}). In order to do this we need the Lagrangian describing the gravitational interaction of a Dp-brane with the target space-time metric, the simplest ansatz is the p-dimensional generalization of the Nambu-Goto action
\be
S=\tau_p\int d\xi^{p+1} \sqrt{-detG^e}.
\label{eq:20}
\ee

Using this amplitude and identifying the string amplitude (fig.\ref{fig4}) with the gravitational field theory amplitude (fig.\ref{fig8}) we get the Dp-brane tension formulae
\be
\tau_p={\pi(4\pi\alpha')^{11-p} \over 256\kappa^2}.
\ee

In principle nothing prevent us from doing the computation of the amplitude in th limit $t\rightarrow 0$. Using again the amplitude for the Dedekind function what we get in this case is
\be
A(y)\simeq (\alpha')^{-d/2}(D-2)\int{dt \over t}t^{-d/2}e^{y^2t/\alpha'}.
\ee
After performing the integration we get,
\be
A(y)\simeq (D-2){1 \over (\alpha')^d}\Gamma\left(-d/2\right )|y|^d.
\label{eq:21}
\ee
We should notice a few things concerning (\ref{eq:21}). First of all the only dependence on the target space-time dimension $D$ is in the irrelevant front factor $(D-2)$. This indicates that (\ref{eq:21}) reflects only the dynamics on the world volume of the Dp-brane. Secondly contrary to the case (\ref{eq:17}) the amplitude (\ref{eq:21}) is singular.

Concerning the amplitude (\ref{eq:21}) we can take the near D-brane field theory limit
\bea
y\rightarrow 0\;,\;\alpha'\rightarrow 0 \;,\; u\equiv {y \over \alpha'}.
\eea
In this case we get:
\be
A(u)\simeq (D-2)\Gamma\left(-d/2\right )|u|^d.
\ee
The limit $\alpha' \rightarrow 0$ of (\ref{eq:17}) can be nicely taken for the special case $p=11$ that correspond to the half dimensional brane in 26-dimensions. Moreover for $p=11$ it is easy to see that the dilaton exchange in the (fig.\ref{fig8}) vanishes. 
\begin{figure}
\begin{center}
\includegraphics[angle=-90,scale=0.4]{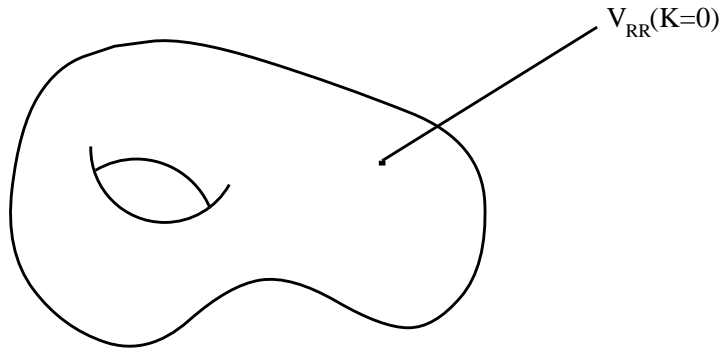}
\caption{RR vertex operator on strigs}
\label{fig9}
\end{center}
\end{figure}

\section{R.R Charged D-branes}


Perhaps the most interesting dynamics of the Dp-branes appears in the case of superstrings (for a good review of superstrings see \cite{gsw}).

The Hilbert space of superstring contains different sectors depending on the world sheet fermion boundary conditions. For periodic boundary conditions, both in the left and right components we have  the so called $R\otimes R$ sector. For antiperiodic boundary conditions we have the $NS\otimes NS$ sector. The two sectors correspond to space-time bosons. In the $NS\otimes NS$ sector we have the standard gravity multiplet containing the dilaton graviton and antisymmetric tensor. In the $R\otimes R$ sector we have also space-time bosons but this time corresponding to the factors appearing in the decomposition of the product of two ``Ramond''-vacua,
\bea
&&\hbox{type A}\;\;8x\bar{8}= [0]+[2]+[4],\nonumber \\
&&\hbox{type A}\;\;8x8=,[1]+[3]+[5]
\label{eq:24}
\eea
with $8$ and $\bar{8}$ representing the different chiralities. Space-time fermions and  gravitons are in the $NS\otimes R$ and $R\otimes NS$ sectors.

One loop modular invariance determines the different GSO projections. At this point we get four different types of superstring theories. For type (A) we get type II with two gravities and no tachyons, also type 0 where we have no gravitons and closed string tachyons. The same for the type B strings.
\begin{figure}
\begin{center}
\includegraphics[angle=-90,scale=0.4]{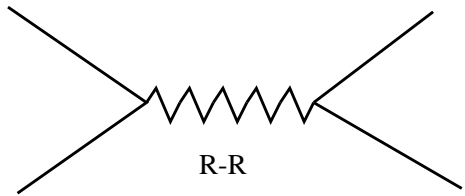}
\caption{RR vertex operator on strigs}
\label{fig10b}
\end{center}
\end{figure}

What is the string meaning of the $R\otimes R$ forms appearing in (\ref{eq:24})?. This is a difficult question from the world sheet point of view. The reason for that comes from the fact that strings are not sources of the $R\otimes R$ fields. The simplest way to see this is to consider a string diagram as the one represented in (fig.\ref{fig9}), where we have a $R\otimes R$ vertex operator inserted in a string amplitude at zero momentum. From the form of the vertex operator we easily see that the amplitude contains a factor of the form
\be
\vec{k}F_{\mu_1...\mu_n},
\label{eq:25}
\ee
for $F$ the $R\otimes R$ stress tension and $k$ the momentum, hence vanishing for $k=0$.

\begin{figure}
\begin{center}
\includegraphics[angle=-90,scale=0.4]{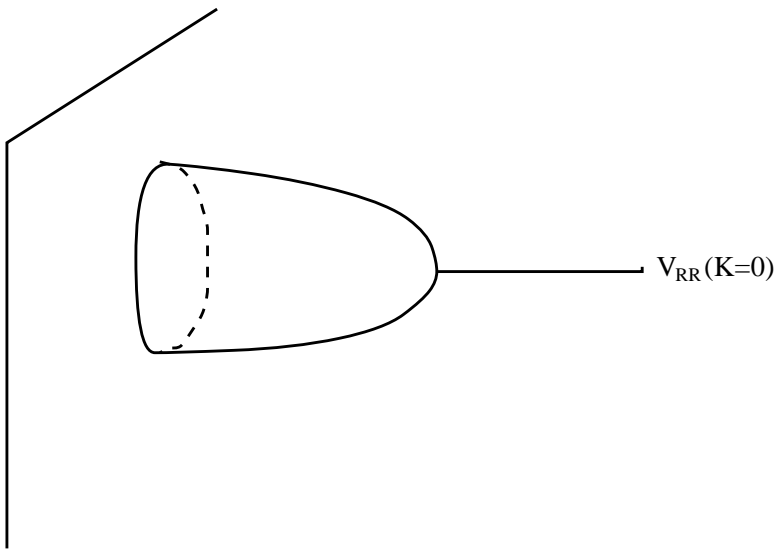}
\caption{RR vertex operator and D-branes}
\label{fig10}
\end{center}
\end{figure}

If strings are not sources of $R\otimes R$ fields is not clear what can be the stringy meaning of defining backgrounds with non vanishing vacuum expectation values for $R\otimes R$ tensors, moreover it is not clear at all what can act as a source of those fields in string theory. The answer to this puzzle comes again from the Dp-branes. In fact Dp-branes are natural sources of $R\otimes R$ charges, the amplitude in (fig.\ref{fig10}) where we represent the amplitude for the Dp-brane  emission of a $R\otimes R$ quanta is now non vanishing mainly due to the fact that the world sheet entering into the game is a disc, due to the Dirichlet boundary condition on the Dp-brane world volume. In order to prove that the amplitude in (fig.\ref{fig10}) is now non vanishing we need to invoke a ``picture changing'' manipulation \cite{fri1} to cancel the $k$ in (\ref{eq:25}).

Once we know that Dp-branes are -or superstrings- sources of $R\otimes R$ fields we can compute the type of interaction between parallel Dp-branes mediated by the interchange of $R\otimes R$ quanta. This amplitude will be exactly the type of amplitude depicted in (fig.\ref{fig4}), but this time we will consider the contribution to the cylinder of $R\otimes R$ states in the spectrum. If we choose a GSO projection implying space-time supersymmetry  we will get exactly the same amplitude with the reverse sing and with $D=10$, the critical dimension for superstrings,
\be
A(y)\sim -(\alpha')^{3-p}\Gamma\left({7-p \over 2}\right )|y|^{p-7}.
\label{eq:26}
\ee
Now we would like to compare this amplitude with the corresponding Feynman diagram in (fig.\ref{fig10b}). In order to do that  we need again to defined  a quantum field theory coupling between the Dp-brane and the $R\otimes R$ $(p+1)$-form
\be
\mu_p\int d\xi^{p+1}A.
\label{eq:27}
\ee
with the kinetic term for the $R\otimes R$ stress tension $F=dA$ given by
\be
\int d^{10}x F².
\ee
By comparing the amplitude in (fig.\ref{fig10})  with space-time (\ref{eq:26}) we get the value of the $R\otimes R$ charge density $\mu_p$ for a Dp-brane:
\be
\mu_p=e^{\Phi}\tau_p\;\;,\;\;\tau_p={\pi(4\pi\alpha')^{3-p} \over \kappa^2}
\ee

Probing the BPS nature of the Dp-brane. The above forms for the interaction of the brane with the $R\otimes R$ field can be deduce from the equation of motion that the open string on the D-brane background must satisfy \cite{lei1}. The resulting action is given by,
\begin{equation}
S = T_p  \int d^{p + 1} \xi
\;e^{-\Phi} \;\sqrt{-\det (g_{\alpha \beta} + b_{\alpha \beta} + 2 \pi \alpha'
f_{\alpha \beta})  },
\label{eq:DBI}
\end{equation}
where $g$, $b$  and $\Phi$ are the pull backs of the 10D metric, antisymmetric tensor and dilaton to the D-brane world volume, while $f$ is the field strength of the world volume $U(1)$ gauge field $A_{\alpha}$ and $T_p=\tau e^{<\Phi>}$.

For the supersymmetric string theory we must extend the action to a supersymmetric Born Infield type action, that includes Chern-Simons type terms that couple the Dp-brane to the $R\otimes R$ fields. Of course this last part of the action, in the leading term correspond to the coupling of equation (\ref{eq:27}).

To simplify the above action we can consider the background space-time to be flat, the Dp-brane almost flat and in the static gauge, hence giving the expansion
\begin{equation}
g_{\alpha \beta} \approx \eta_{\alpha \beta} + \partial_\alpha X^a
\partial_\beta X^a+{\cal O} \left((\partial X)^4 \right).
\end{equation}
On the top of this we can consider vanishing antisymmetric field $b$ and that $2 \pi \alpha' F_{\alpha \beta}$ and $\partial_\alpha X^a$ are
small and of the same order. The resulting low energy action is:
\bea
&&S =\tau_p\int d\xi^{p+1} \sqrt{-det\hat{G}}+\nonumber \\
&&\frac{1}{4g_{{\rm YM}}^2} \int d^{p + 1} \xi \left(F_{\alpha \beta} 
F^{\alpha \beta} +\frac{2}{(2 \pi \alpha')^2} \partial_\alpha X^a 
\partial^\alpha X^a\right),
\eea
where the Yang-Mills coupling is given by,
\begin{equation}
g_{YM}^2 = \frac{1}{4 \pi^2 \alpha'^2 \tau_p} 
= \frac{g}{\sqrt{\alpha'}}  (2 \pi \sqrt{\alpha'})^{p-2}.
\label{eq:39}
\end{equation}
The second term correspond to the dimensional reduction of the ten dimension SYM action. Once the fermions are include, the resulting action is the dimensional reduction of supersymmetric $N=1$, $U(1)$ Yang-Mills theory in ten dimension.
\begin{equation}
S = \frac{1}{g_{YM}^2}  \int d^{10}\xi \; \left(
 -\frac{1}{4} F_{\mu \nu}F^{\mu \nu} + \frac{i}{2}  \bar{\psi} \Gamma^\mu 
\partial_{\mu} \psi \right).
\end{equation}


\section{Microscopic Dp-brane String Amplitudes and Metrics}


Before going into the supergravity effective Lagrangian associated with superstring, let us motivate from the previous discussion on amplitudes the form of the metric generated by a Dp-brane in the limit $t\rightarrow 0$ i.e.
\be
A(y)\sim (\alpha')^{3-p}\Gamma\left({7-p \over 2}\right )|y|^{p-7}.
\label{eq:30}
\ee

In the post-Newtonian approximation we can think of (\ref{eq:30}) as a gravitational correction to the flat metric. Moreover we can invoke the T-duality relation between different Dp-branes to look for an ansatz for the metric of the type
\be
ds^2=\left(1-{g(\alpha')^{7-p/2}\Gamma\left({7-p \over 2}\right ) \over y^{7-p}}\right)dx_{||}^2+\left(1+{g(\alpha')^{7-p/2}\Gamma\left({7-p \over 2}\right ) \over y^{7-p}}\right)dx_{\perp}^2.
\label{eq:31}
\ee

In the next section we will see that (\ref{eq:31}) is at first order in the string coupling constant $g$ a solution to the supergravity equations.


\section{Supergravity as an Effective Theory}


In this section we are going to talk about the supergravity description of Dp-branes. This kind of description involves the effective theory describing the low energy massless states of the Type II superstring theories, where these massless states are the graviton, the dilaton the NS two-form and the corresponding $R\otimes R$ forms. An important point to consider about this effective theories, is that their specific form is completely determined by requiring supersymmetry on the massless spectrum just mentioned. Therefore if the superstring theory has a given symmetry, the effective theory should also be invariant under this symmetry (supersymmetry acts as a protecting shield to quantum corrections). Actually the action itself doesn't have to be invariant under this symmetry, is good enough to have invariant field equations. Recall that what we really obtain from the beta function in the non-linear supersymmetric $\sigma$-model is the field equation for some of the massless modes. The other important point that we should keep in mind is that because the Dp-brane are BPS states, some of their characteristics are also protected from quantum effects, for example the mass, charge and even the degeneracy of their spectrum, as we move on the moduli space of the full superstring theory. Therefore the low energy description of Dp-branes and BPS objects in general is very important. In fact these solutions are one of the most powerful ways to study the new dualities within the different string theories.  

The effective theories (for a good review see \cite{kel1}) obtained from the type II superstrings are the type IIA and type IIB supergravity. Type IIA has for spectrum the graviyton $G$, dilaton $\Phi$, the NS two-form $B$, and in the Ramod sector the one-form potential $A_{[1]}$ and the three-form potential $A_{[3]}$. The action in the string frame is given by 
\begin{eqnarray}
S_{IIA} &= {1\over 2\kappa^2}\int\! d^{10}x \bigg\{ \sqrt{-G}\,
e^{-2\Phi}\big[ R + 4|d\Phi|^2 -{1\over 12} |H|^2\big] \cr
&- \sqrt{-G}\, \big[{1\over 4}|F_{[2]}|^2 + {1\over 48}|F_{[4]}|^2\big] \bigg\}\  + {1\over 4\kappa^2}\int F_{[4]}\wedge F_{[4]}\wedge B \, ,
\end{eqnarray}
where $F_{[4]}= dA_{[3]} + 12B \wedge F_{[2]}$ is the non-linear version of the 4-form field
strength.

The type IIB spectrum is given by the graviton $G$, dilaton $\Phi$, the NS two-form $B$ and in the Ramond sector we get a pseudoscalar potential $A_{[0]}$, a two-form potential $A_{[2]}$ and a four-form potential $A_{[4]}$ whose field strength is self-dual. The action in the string frame is given by
\begin{eqnarray}
\label{eq:}
S_{IIB} &= {1\over 2\kappa^2}\int d^{10}x {\sqrt {-G}}
\bigg\{e^{-2\Phi}\big[ R + 4|d\Phi|^2-{1\over 12}|H|^2\big]  -2|d\ell|^2  \cr
&-{1\over 3}|F_{[3]}-\ell H|^2  
 - {1\over 60}|F_{[5]}|^2\bigg\} -{1\over 4\kappa^2}\int A_{[4]}\wedge H\wedge F_{[3]},
\end{eqnarray}
where we have ignore the self-duality condition of $F_{[5]}$ to write this action\footnote{ The self-duality condition for the $F_{[5]}$ makes very difficult to write an action for type IIB supergravity. In any case we can always introduce this condition as a constraint at the level of field equations}. The full non-linear Bianchi identity satisfied by $F_{[5]}$ is now $dF_{[5]} = H\wedge F_{[3]}$. By combining this `modified' Bianchi identity with the self-duality condition on $F_{[5]}$ we deduce that $d\star F_{[5]} = H\wedge F_{[3]}$, which is just the $A_{[4]}$ field equation. Thus, the modification of the Bianchi identity is needed for consistency  with the self-duality condition.

To study the Dp-brane solutions of the above actions it is convenient to consider truncations on the Lagrangian that simplify and clarify the task of finding out the right ansatz. Basically we can always set to zero all but one of the field strength, living us we the following type of action (written in the Einstein frame for simplification)
\be
I=\int dx^{10}\sqrt{-G}\Big[R-\ft12|d\Phi|^2-\ft{1}{2(d+1)!}e^{a\Phi}|F|^2_{[d+1]}\Big].
\label{eq:36}
\ee
where $a^2=(d-4/2)^2$.

With this Lagrangian we are ready to look for solutions that are sensible to be interpreted as Dp-branes. The ansatz we have to consider in the most simplified case is:
\begin{itemize}
\item Asymptotic flat space time.
\item Broken Lorentz invariance from $SO(1,9)$ to $SO(1,p)\otimes SO(9-p)$.
\item A bosonic supersymmetric solution.
\end{itemize}
This conditions can be understand as follows: The first item allows us to define the meaningful quantities that characterize a BPS state like the mass, charge, and supersimetric killing spinors. The second item assumes we are considering the so called static gauge on the coordinates used i.e. the world-volume coordinates of the corresponding Dp-brane are exactly the same as the first $d=p+1$ space-time coordinates $x^\mu$. Also implies that none of the configuration depends on the above coordinates and that, in the simplest assumption all the possible non-trivial dependence of the involved fields goes in term of a radial coordinate defined on the perpendicular space to the brane, with coordinates $y^m$. The resulting form of the ansatz is given by
\be
ds^2 = e^{2A(r)}dx^\mu dx^\nu\eta_{\mu\nu} +
e^{2B(r)}dy^mdy^n\delta_{mn}\ ,
\ee
where $r=\sqrt{y^my^m}$.
For the dilaton we have only radial dependence $\Phi=\Phi(r)$ and for the RR field we have two possibilities, one related to the electric solution and the other to the magnetic solution, the form of the ansatz for the electric case is
\be
F_{d+1}=dA_{d}\;\;;\;\;\;A_{\mu_{1}.......\mu_{d}}=\epsilon_{\mu_{1}.......\mu{d}}e^{C(r)}.
\ee
For the magnetic ansatz we can only give an expression in terms a o the strength filed, as there is no global definition for the associated potential,
\be
F^{\rm (mag)}_{m_1\cdots m_n} = g_{\bar{d}}\epsilon_{m_1\cdots
m_np}{y^p\over r^{n+1}}\ ,\hspace{.5cm}
\mbox{others zero}.
\ee

By now the only condition we haven't used is the supersymmetry character of the solutions. In forthcoming sections we will be dealing more carefully with the supersymmetric equations, but for our actual proposes, it is sufficient to know that the constraints imposed by supersymmetry restrict the number of independent functions from four $(A,B,\Phi,C)$ to one, let say $C$. Therefore using the field equations obtained varying the action (\ref{eq:36}), we get the electrically charged Dp-brane solution (in the Einstein frame),
\begin{eqnarray} &&ds^{2}=H^{\frac{-\bar{d}}{d+\bar{d}}}dx^{\mu}dx_{\mu}+
H^{\frac{d}{d+\bar{d}}}dy^{m}dy_{m} \nonumber  \\ 
&&e^{\Phi}=H^{\frac{ a}{2}} \\
&&H(r)=\left\{ \begin{array}{ll}
	1+\frac{K_{d}}{r^{\bar{d}}}\;\;\;  \mbox{if $\bar{d}>0$} \nonumber\\
	C_{o}+K_{o}ln(r)\;\;\;\mbox{if $\bar{d}=0$}
	\end{array}
	\right. 
\end{eqnarray}
where $a=-\sqrt{4-\frac{2d\bar{d}}{d+\bar{d}}}$.

Note that the constant integration $K_d,K_o,C_o$ and the parameter $g_{\bar{d}}$ are not fixed by this ansatz. Also we have introduce a new constant $\bar{d}$, which satisfies the equation $d+\bar{d}=8$, being the worldvolume dimension of the magnetic solution associated to a strength field $F_{[d+1]}$. To get the solitonic solutions, one replaces $d$ by $\bar{d}$ and set $a(\bar{d})=-a(d)$. This will give a $\bar{d}$-brane magnetically charged.

We can define the mass, electric charge and magnetic charge of the above solutions, by the expressions,
\bea
m&=& {1 \over 2\kappa}\int_{S^{\bar{d}+1}} d^{\bar{d}}{\Sigma^m(\partial^nh_{mn}-\partial_mh_n^n)}, \nonumber \\
e_{d} &=& {1 \over \sqrt{2}\kappa}\int_{S^{\bar{d}+1}} e^{-a(d)\phi} F^{*},  \\
g_{\bar{d}} &=& {1 \over \sqrt{2}\kappa}\int_{S^{d+1}} F, \nonumber
\eea
where we expand the metric $G$ as $G=\eta + h$. When we solve the mass, and charges for the above metrics we find that this solutions have the characteristic relation between the mass and charge. Also the electrically charge solution and the magnetically charge solution satisfy the Dirac quantization condition
\be
e_{d} g_{\bar{d}}=2\pi n,
\ee
with $n$ an integer.

A very peculiar fact about the electrically charge Dp-branes is that they solve the supergravity field equations with sources at the origin, where the source is the action of the elementary Dp-brane (\ref{eq:20}) plus the coupling with the RR potential. This fact allow us to define the value of $K_,e_d,g_{\bar{d}}$ in terms of the string coupling constant and the string length $\alpha'$, giving $K_d=2^{6-d}\pi^{(10-d)/2}\Gamma(6-d/2)gl_p^{d-4}$. For future discussions it is important to rewrite the Dp-brane metrics in the string frame, here we show the result:

For the elementary case, we have

\begin{eqnarray}
&&ds^{2}=H^{-1/2}dx^{\mu}dx_{\mu}+ H^{1/2}dy^{m}dy_{m} \nonumber \\
&&e^{\Phi}=H^{1-d/4}  \nonumber \\
&&H(r)=\left\{ \begin{array}{ll}
	1+\frac{K_{d}}{r^{\bar{d}}}\;\;\;  \mbox{if $\bar{d}>0$} \\
	C_{o}+K_{o}ln(r)\;\;\;\mbox{if $\bar{d}=0$}
	\end{array}
	\right. \\
&&F_{\mu_{1}.......\mu_{d}m}=\epsilon_{\mu_{1}.......\mu_{d}}\partial_{m}H^{-1} \nonumber
\end{eqnarray}
where $\mu=0,...,d-1$, $\;\;m=d,...,9$.

\vspace{12pt} 
For the the solitonic case,

\begin{eqnarray}
&&ds^{2}=H^{-1/2}dx^{\mu}dx_{\mu}+ H^{1/2}dy^{m}dy_{m}  \nonumber \\
&&e^{\Phi}=H^{d/4-1} \nonumber \\
&&H(r)=\left\{ \begin{array}{ll}
	1+\frac{K_{\bar{d}}}{r^{d}}\;\;\;  \mbox{if $d>0$}\\
	C_{o}+K_{o}ln(r)\;\;\;\mbox{if $d=0$}
	\end{array}
	\right. \\
&&F_{m_{1}.......m_{d+1}}=g_{\bar{d}}\epsilon_{m_{1}.......m_{d+1}l}\frac{y^{l}}{r^{d+2}}. \nonumber
\end{eqnarray} 
where $\mu=0,...\bar{d}-1$, $\;\;m=\bar{d},...,9$.

The Lagrangian of equation (\ref{eq:36}), as we just showed above, contains the class of solutions to its field equations, relevant for the Dp-brane studies. This type of Lagrangian is not restricted to the ten dimensional bosonic sector of a truncated type II supergravity, in fact we found the same structure in the bosonic sector of eleventh dimensional $N=1$ supergravity theory. Where the  bosonic part of the Lagrangian is given by
\begin{eqnarray}
S &=& {1\over 2\kappa^2}\int \! d^{11}x\, \sqrt{-\hat{G}}[R - {1\over 48}|\hat{F}|^2]  +\ {1\over 12\kappa^2}\int \hat{F} \wedge \hat{F} \wedge \hat{A}.
\end{eqnarray}

This time the field potential is a three-form $\hat{A}_{[3]}$, and we put hats on each field to difierenciate from ten dimensional variables. Therefore the associated p-branes solutions are the electrically charged M3-brane and the magnetic M5-brane. Their names comes from the idea that $D=11$, $N=1$ supergravity is the low energy effective theory of the famous M-theory. In fact $D=11$, $N=1$ supergravity compactified on a small circle gives type IIA supergravity in ten dimensions, plus the relevant RR field strength, coming as Kalusa Klein modes on the dimensional reduction. This is part of the conjectured relation of type IIA superstring theory and M-theory. In any case the M-branes solutions of $D=11$, $N=1$ supergravity  are given below. For the electric M3-brane we have,
\begin{eqnarray}
&&ds^{2}=\left(1+{K_3\over r^6}\right)^{-2/3}dx^{\mu}dx_{\mu}+ \left(1+{K_3\over r^6}\right)^{-1/3}dy^{m}dy_{m}, \nonumber \\
&&A_{\mu\nu\rho} = \epsilon_{\mu\nu\rho}\left(1+{K_3\over r^6}\right)^{-1},
\end{eqnarray}
where $\mu=0,...,2$, $\;\;m=3,...,10$. For the magnetic M5-brane we get
\begin{eqnarray}
&&ds^{3}=\left(1+{K_6\over r^3}\right)^{-1/3}dx^{\mu}dx_{\mu}+ \left(1+{K_6\over r^3}\right)^{2/3}dy^{m}dy_{m}, \nonumber \\
&&F_{mnop} = 3K_6\epsilon_{mnopq}{y^q\over r^5},
\end{eqnarray}
where $\mu=0,...,5$, $\;\;m=6,...,10$.

As in the case of the Dp-branes the constant of integration can be related to the relevant structure constant of M-theory, obtaining that $K_6=\pi l_p^3$ and $K_3=2^2\pi^2l_p^6$.
\begin{figure}
\begin{center}
\includegraphics[angle=-90,scale=0.4]{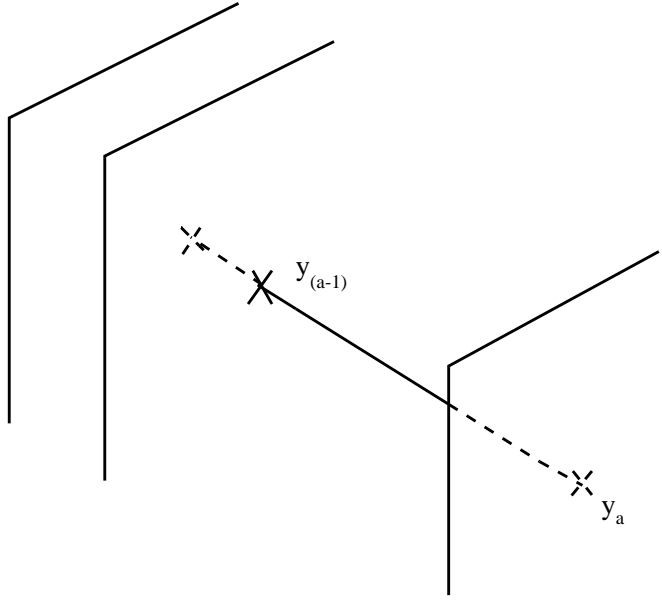}
\caption{Multicentre D-branes}
\label{fig11}
\end{center}
\end{figure}

Before finishing the discussion of Dp-branes as solutions of the effective low energy theories of superstrings theories, we should recall that there is a very simple generalization to the above solutions which will be used extensible on future discussions. Basically we can consider what is called a multiple centre branes solution i.e. a solutions representing a given number of branes, say $N$ in a parallel configuration. Because the branes are all oriented in the same way, the supersymmetry conserved is the same as in the previous cases. If we prepare the solution such that each brane is located at $\vec{y}_a$ (see fig. \ref{fig11}) then the only difference in the brane solution comes in the form of the harmonic function $H$
\be
H=1+\sum_{a=1}^N{K_{(d,a)} \over |\vec{y}-\vec{y}_a|^{\bar{d}}}.
\ee
The form of the electric field ansatz is unchanged , but the magnetic field needs the following change,
\be
F_{m_{1}.......m_{d+1}}={-1 \over \bar{d}}\epsilon_{m_{1}.......m_{d+1}}\partial_p\sum_{a=1}^N {g_{(\bar{d},a)} \over |\vec{y}-\vec{y}_a|^{\bar{d}}}.
\ee
This solutions are relevant when we consider Dp-brane configurations with open string sectors including Chan-Paton factors. In principle this factors produces a family of Dp-branes, rather than a single brane. In particular strings stretching between different branes correspond to massive states of the gauge theory defined on the world volume of the Dp-branes. If we start with $N$ different Dp-branes all at the same place, we have a $SU(N)$ $(U(N))$ gauge theory living on the world volume of the brane. By pulling out one of these branes, we break the group  $SU(N)$ to $SU(N)\otimes U(1)$, and the mass of the associated W-vector boson is given in terms of the distance between the bunch of $N-1$ Dp-branes and the pulled out Dp-brane, $r$ and the squared string length 
\be
|mass|={r \over \alpha'}.
\ee
Therefore we have a very nice geometrical picture of symmetry breaking on gauge theories.

As a last remark we introduce the notion of T-duality for the RR-fields. From the point of view of the string theory the T-duality transformation is (as we said on section 2) a hybrid parity operation, this parity restricted to the anti-holomorphic worlsheet sector is realized in the spinor space as the operator $-i\Gamma^9\Gamma_{11}$ (where $x^9$ is the compact direction). As the RR-fields are bispinors defined by the corresponding vertex operator, its transformation is defided as,
\be
F=-iF\Gamma^9\Gamma_{11}.
\ee
therefore we get the final relation in terms of the index of $F$ as,
\bea
F_{\mu_1...\mu_n}=-F_{9\mu_1...\mu_n} \nonumber \\
F_{9\mu_1...\mu_n}=-F_{\mu_1...\mu_n}
\eea
The general form of the transformation in the case of non trivial  background metric NSNS-antisimmetric field and dilaton can be found on \cite{berg1}.
 

\section{Near horizon Geometry}


Given a Dp-brane metric
\be
ds^{2}=H^{-1/2}dx^{\mu}dx_{\mu}+ H^{1/2}dy^{m}dy_{m}
\ee
with 
\be
H=1+{(4\pi)^{(5-p)/2}(\alpha')^{(7-p)/2}gN\Gamma((7-p)/2) \over r^{7-p}}
\label{eq:60}
\ee
where $r=\sqrt{y^{m}y_{m}}$, the near horizon geometry is defined by the double limit \cite{mal1}
\bea
\al'\rightarrow 0\;\;&,&\;\;r\rightarrow 0 \nonumber \\
{r \over \alpha'}&=&U
\label{eq:61}
\eea
with $U$ arbitrary. Using (\ref{eq:61}) we can rewrite the harmonic funtion $H$ as
\be
H=1+{(4/pi)^{(5-p)/2}gN\Gamma((7-p)/2) \over U^{7-p}(\alpha')^{(7-p)/2}} \approx {(4/pi)^{(5-p)/2}gN\Gamma((7-p)/2) \over U^{7-p(\alpha')^{(7-p)/2}}}
\label{eq:62}
\ee
Geometrically we can think of (\ref{eq:60}) as a function defined on the $(\alpha',r)$ plane. In these conditions (\ref{eq:61}) defines a blow up of the point $(0,0)$ generating a divisor with coordinate $U$.
Using (\ref{eq:62}) we get for the D3-brane the following near horizon geometry,
\bea
ds_{3}^2&=& \alpha'\left[ \frac{U^2}{(gN4\pi)^{1/2}}dx_{||}^2+\frac{(gN4\pi)^{1/2}}{U^2}dU^2 +(gN4\pi)^{1/2}d\Omega_{5})\right] \nonumber \\
e^\phi&=&g.
\label{eq:63}
\eea
which is the metric of $AdS_5\otimes S^5$.

For a generic Dp-brane we can write the metric in a similar form provided we use instead of $g$ -the string coupling constant- the Yang Mills coupling constant on the Dp-brane which is given by (\ref{eq:39}).
\begin{equation}
g_{YM}^2 = \frac{1}{4 \pi^2 \alpha'^2 \tau_p} 
= \frac{g}{\sqrt{\alpha'}}  (2 \pi \sqrt{\alpha'})^{p-2}.
\end{equation}
Note that the Yang-Mills coupling constant is also defined as the result of a limiting procedure where $g$ is sent to infinite or zero, depending on the value of $p$. Using the above equation, the near horizon form of the harmonic function $H$ becomes
\be
H={d_pg_{YM}^2N \over U^{7-p}(\alpha')^{-2}}
\label{eq:65}
\ee
where $d_p= 2^{7-2p}\pi^{(9-3p)/2}\Gamma((7-p)/2)$. Therefore we get the following near horizon metric,
\bea
ds^2&=& \alpha'\left[ \frac{U^{(7-p)/2}}{g_{YM}(Nd_p)^{1/2}}dx_{||}^2+
\frac{g_{YM}(Nd_p)^{1/2}}{U^{(7-p)/2}}dU^2 +g_{YM}(Nd_p)^{1/2}U^{p-3}d\Omega_{5})\right] \nonumber \\
e^\phi&=&2\pi^{2-p}g_{YM}^2\left(\frac{g_{YM}^2Nd_p}{U^{7-p}}\right)^{(3-p)/4}.
\label{eq:67}
\eea

How should we interpret the near horizon geometries? (To find a more complete list of reference and a deeper introduction see \cite{mal2}). Since we are performing the $\alpha'\rightarrow 0$ limit we can think of these geometries as related somehow to the pure Yang-Mills theory living on the Dp-brane. We can try to establish this relation from 3 different points of view.

\begin{enumerate}
\item {\bf Gauge Singlets}: The main idea underlying this approach will consist in looking for an isomorphism between gauge singlets of the gauge theory on the Dp-brane and the spectrum of supergravity defined on the near horizon geometry. Since supergravity is not a complete theory we should look for its string ancestor on the near horizon geometry background. Supergravity approximation will be good if the curvature of the near horizon geometry is large in string units, which in general will implies to take a large $N$ number of branes.

Notice that the standard open-closed relation in string theory always allows to think of the closed gravitational sector in terms of gauge singlets of the open string theory. The problem is to find a limit where the closed string spectrum precisely describes the physics of the gauge singlets (glue-balls) of the field theory limit $\alpha'\rightarrow 0$ of the open string sector.

\item {\bf Renormalization Group}: In this approach we start by identifying the blow up variable $U$ with the renormalization group scale of the gauge theory. The direct connection between string theory on the near horizon geometry and the pure Yang-Mills theory on the Dp-brane in the  $\alpha'\rightarrow 0$ limit, will be identifying the renormalization group equation of the field theory with the dilaton behaviour dictated by the Dp-brane near horizon geometry through the string beta function equations.

\item {\bf Matching of Symmetries}: A more concrete procedure to find a relation between the near horizon geometry and the pure Yang-Mills theory  is by matching the symmetries of the Yang-Mills theory with the isometries of the near horizon geometry. the relevant data for the Yang-Mills theory are their supersymmetries and global R-symmetries. From the supergravity point of view we should consider the corresponding superalgabra.
\end{enumerate}

Let us consider the simplest case of the D3-brane. The corresponding super-Yang-Mills theory is $D=4,N=4$. This is a theory with $16$ supersymmetries and conformal invariance that is described by the superconformal $N=4$ superalgebra. As we will see in the next section, the matching with the near horizon symmetries comes from the fact that in $AdS_5\otimes S^5$ we get an enhancement of supersymmetry.

From the point of view of the renormalization group the matching is in this case trivial since on one side we have $\beta=0$ and on the other a constant dilaton.

The most difficult aspect is of course the correspondence between gauge singlest and supergravity fields. Here the geometry of $AdS$ space time is specially important. In fact we will look for a correspondence between supergravity fields $\psi_i$ and gauge singlets observable $\theta_i$ of super-Yang-Mills theory is $D=4,N=4$ in such a way that $\psi_i$ at the boundary of $AdS_5$ can act as a source for the operators $\theta_i$ through an interaction term of the type,
\be
\int{dx^4\;\theta_i(x)\psi_i(x)}
\ee
with $\psi_i(x,U)$ the boundary value of $\psi_i(x,U)$.

The operator $\theta_i$ can be characterized by their conformal weight $\Delta_i$ hence we need $\psi_i$ to have dimensions of the type $[lenght]^{\Delta_i-4}$. This condition fixes the mass of the field  $\psi_i$ to be determined by the following relation,
\be
\Delta_i=2+\sqrt{1+R^2m^2}
\ee
for $R$ the $AdS_5$ radius. Once we have this correspondence between $AdS_5$ fields $\psi_i$ and the observable $\theta_i$ of super-Yang-Mills theory is $D=4,N=4$ the recipe for computation of the amplitudes is given by,
\be
\langle e^{\int{dx^4\;\theta_i(x)\hat{\psi}_i(x)}}\rangle_{SSYM}=e^{S_{sugra}(\psi_i(x,U))}
\label{eq:70}
\ee
with $\psi_i(x,U)\mid_{U=0}=\hat{\psi}_i$. Relation (\ref{eq:70}) is of course valid as long as the $AdS_5$ radius is large enough. It is conjectured that string corrections to the r.h.s. of equation (\ref{eq:70}) still reproduce SSYM dynamics, but for the time being this conjecture has not been proved.

In Summary the correspondence between $AdS_5\otimes S^5$ supergravity and super-Yang-Mills theory is $D=4,N=4$ is based on two facts: i) the isomorphism of superalgebras for $N=4$ SSYM and $N=8$ $AdS_5$ supergravity, isomorphism that actually depends on the enhancement of supersymmetry in $AdS_5$ and ii) the special structure of the conformal infinity of $AdS_5$, which is isomorphic to four dimensional Minkowski space time. How to extend this picture to non-conformal cases is an interesting open problem. Most likely the starting point in trying to solve this problem will consist in a systematic string reinterpretation of quantum field theory renormalization group equations.


\section{Supersymmetry}


M-branes and Dp-branes are known to be BPS states. That is to say that saturate a Bogoumoly inequality relating their mass and charge. In turns this imply that the supersymmetry preserved by this objects is a fraction of the maximal supersymmetry appearing on the theory. In this case we have only $16$ real supercharges or $1/2$ of the maximal number of real supercharges that is $32$ for $D=11$ , $N=1$ supergravity and  $D=10$ , $N=2$ supergravity. A nice way to see why we have this relation between BPS states and fractional number of the maximal supersymmetry conserved, is to look  on the superalgebra of the above theories at infinity (where we can defined mass and charge, associated with the Poincare group). The general form of this superalgebra is, in the presence of branes given skematically by
\be
\{ Q,Q \}=\Gamma^M P_m+\Gamma ^{M_0...M_p}Z_{M_0...M_p},
\ee
where $P$ is the momentum and $Z$ is form defining the charge carried by the brane. The BPS status of the brane tell us that the mass and the charge are equal, therefore in  static configurations we get
\be
\{ Q,Q \}=m(1\pm\Gamma ^{0...p}),
\label{eq:52}
\ee
clearly the eigenspinors of $\{ Q,Q \}$ satisfy
\be
\Gamma ^{0...p}\epsilon^\pm_0=\pm\epsilon^\mp_0.
\label{eq:53}
\ee
The operator $\Gamma ^{0...p}$ squares to the identity and is trassless. Therefore the number of independent eigenvalues is one half of the maximum possible, in other words $16$.

We can see the above mechanism in the supergravity solutions of the Branes that we study before. In the type II supergravity case the supersymmetry transformation acting on the fermionic field gives the following equations,
\begin{equation}
\delta \psi_{M} = \partial_{M} \epsilon - {1\over 4}
\omega_{M} ^{AB}
\gamma_{AB}
\epsilon + {(-1)^p \over 8(p+2)!} e^{\phi} F_{M_1 ... M{p+2}}
\gamma^{M_1 ...
M_{p+2}} \gamma_{M} \epsilon^{'} _{(p)}
\end{equation}
\begin{equation}
\delta \lambda = \gamma^{M} (\partial_{M} \phi) \epsilon + {3-p\over
4(p+2)!}
e^{\phi} F_{M_1 ... M_{p+2}} \gamma^{M_1 ... M_{p+2}}
\epsilon^{'}_{(p)}
\end{equation}
\begin{equation}
\epsilon_{(0,4.8)} ^{'} = \epsilon
\qquad
\epsilon_{(2,6)} ^{'} = \gamma_{11} \epsilon
\qquad
\epsilon_{(-1,3,7)} ^{'} = \imath \epsilon
\qquad
\epsilon_{(1,5)} ^{'} = \imath \epsilon^{\ast}
\end{equation}
where $\epsilon $ is a 32-component spinor, and $\omega$ is the spin
connection
After solving for the Dp-brane solutions we found the equations
\bea
\delta \lambda &=& H^{-1/4}\gamma^r\partial_r\phi\epsilon+{(3-p)e^\phi(\partial_r H)\gamma^r \over 4H^{8-p\over 4}}
\gamma_0 ... \gamma_p \epsilon^{'}  \nonumber \\
\delta \psi_{\alpha} &=& \partial_{\alpha} \epsilon + {(\partial_r H)\over 8H^{3\over 2}}\gamma^r \gamma_{\alpha}\epsilon + {e^\phi(\partial_r H)\over 8H^{9-p\over 4}}\gamma^r \gamma_{\alpha}\gamma_0 ... \gamma_p \epsilon^{'} \nonumber \\
\delta \psi_r &=& \partial_r \epsilon -{e^\phi(\partial_r H) \over 8H^{7-p \over 4} } \gamma_0 ...\gamma_p \epsilon ^ {'} \nonumber
\label{susy1}
\eea
\begin{equation}
\epsilon_{(0,4.8)} ^{'} = \epsilon 
\qquad
\epsilon_{(2,6)} ^{'} = \gamma_{11} \epsilon
\qquad
\epsilon_{(-1,3,7)} ^{'} = \imath \epsilon
\qquad
\epsilon_{(1,5)} ^{'} = \imath \epsilon^{\ast}
\label{eq:57}
\end{equation}
Where we have used the split $M=(\alpha,r,\theta)$ where  $(r,\theta)$ are perpendicular coordinates to the brane, also $\epsilon $ is a 32-component spinor, and $\omega$ is the spin connection \cite{ber1}. Note that we have on propose leave the dilaton unspecified in terms of $H$.

The first thing to note about this system of equation as is that for the D3-brane case the dilatino equation is satisfied independently of the type of spinor consider, therefore we are left with only the gravitino constraints \footnote{This case is similar to the M2-brane and the M5-branes, where the supersymmetry equations are written in terms of the gravitino constraints only (in $D=11$ there is no dilatino)}. This is a consequence of the fact that in this case the dilaton is constant. For the other Dp-branes the dilaton equation is present and the solutions are dilatonic.

The dilatino equation is up to multiplicative factors the projector operator appearing on equation (\ref{eq:52}), therefore its presence implies the breaking of $1/2$ of supersymmetry. Also note that the gravitino equation on the worldvolume coordinates is proportional to the same projector up to a additive factor of the form $\partial_\alpha\epsilon$. The last equation referring to the gravitino components on the perpendicular space doesn't correspond to to the full projector but just the part showed on equation (\ref{eq:53}) plus a partial derivative on the perpendicular direction. The solution of this equations is obtained by assuming no dependence on the worldvolume coordinates plus using the eigenspinors defined on equation (\ref{eq:53}), giving the result
\be
\epsilon=H^{-1/8}\epsilon_0.
\ee
where $\epsilon_0$ satisfied that $\epsilon_0=- \gamma_0 ...\gamma_p \epsilon ^ {'}$. Hence we get only the expected $16$ real supercharges.

Let`s next consider the case for the near horizon geometries obtained from the Dp-branes solutions. At first this question could be consider a bit trivial, after all we are talking about Dp-brane solutions in certain regions, and we already know the BPS structure of this type of solutions. Nevertheless note that the near horizon limit change the asymptotic structure of the metrics. In those cases we lost the Minkowski structure and the relation with the Poincare invariants like the mass. The argument showed at the beginning of this section simply doesn't applied to this situation.

To study the supersymmetry properties of these near horizon geometries we start with the equation (\ref{eq:57}). Our goal is to define the near horizon limit of this set of equations. For the gravitino equations, the limiting recepee is quite simple due to the fact that the gravitino supersymmetry variations are components of a one-form, therefore we have an geometrical object where to define the near horizon limit, basically
\bea
\delta \hat{\psi}\equiv lim_{NH} \left(\delta\psi\right)=
lim_{NH}\left(dx^M\delta \psi_M\right)=
lim_{NH}\left(dx^M\right)lim_{NH}\left(\delta \psi_M\right), \nonumber \\
\eea
hence we get
\bea
\delta \psi_{\alpha} &=& \partial_{\alpha} \epsilon + {(\partial_u h)\over 8h^{3\over 2}}\gamma^u \gamma_{\alpha} [\epsilon + \gamma_0 ... \gamma_p \epsilon^{'}], \nonumber \\
\delta \psi_u &=& \partial_u \epsilon -\left({\partial_u h \over 8h}\right)
\gamma_0 ...\gamma_p \epsilon ^ {'},  
\eea
where $h=g^2_{YM}/U^{7-p}$.
On the other hand the dilatino equation is a bit more subtle, as it is a scalar from the point of view of ten dimensional supergravity. We can define its near horizon limit by consider the dilatino variation as part of the relevant gravitino in a higher dimensional theory like eleventh  or twelve dimensional dimensional supergravity.  Once this is done the resulting near horizon limit is,

\bea
\delta \lambda &=& \alpha'^{1/6}\left[{(3-p)(\partial_u h)\gamma^u \over 4h^{5\over 4}}[\epsilon + \gamma_0 ... \gamma_p \epsilon^{'}]\right]. 
\eea

In this case the dilatino equation goes to zero in the near horizon limit, giving no constraints. Nevertheless the consistency conditions corresponding to the gravitino equation, contains the dilatino equation among others. After the usual decomposition of the killing spinors, defined by the canonical projector on this ansatz, we find that one half of the supersymmetry is always preserved if the dilatino constraint is satisfied , but the other half is conserved only if the function $h$, behaves as
\be
h \propto {1 \over U^4},
\ee
therefore we found the possibility of enhancement for $p=3$ only. The killing spinors equation found in this case corresponds to the AdS equation
\bea
\partial_{\alpha} \epsilon -\frac{u}{g^{1/2}_{YM}}\gamma_{\alpha}(1 -\gamma_u)\epsilon =0, \nonumber \\
\partial_u \epsilon -\frac{1}{2u}\gamma_u\epsilon=0,  
\eea
with the well known solution. 
\bea
\epsilon_1&=&u^{1/2}\epsilon^+_0, \nonumber \\
\epsilon_2&=&\left(u^{-1/2}+\frac{u^{1/2}}{2g^{1/2}_{YM}}x^\alpha\gamma_\alpha\right)\epsilon^-_0.
\eea

The M-theoretical cases go along the same line of reasoning as before, here we show in detail the M2-brane only. The supersymmetry transformation for the gravitino in eleventh dimensions give,
\begin{equation}
\delta \hat{\psi}_{M} = \partial_{M} \hat{\epsilon} - {1\over 4}
\hat{\omega}_{M} ^{AB}\hat{\gamma}_{AB}\hat{\epsilon} - {1 \over 288}\left ( \hat{\gamma}_{M}^{BCDE}-8\hat{e}_{M}^B\hat{\gamma}^{CDE}\right)\hat{F}_{BCDE},
\end{equation}
where $M,N,O$ refers to Einstein index and $A,B,C$ refers to Lorenz index, $\hat{e}$ is the elfvien and $\hat{\omega}_{M} ^{AB}$ is the spin connection. Also we have hatted the variables to avoid confusion with ten dimensional variables. After solving for the M2-brane ansatz we get,
\bea
\delta \hat{\psi}_{\alpha} &=& \partial_{\alpha} \hat{\epsilon} - {(\partial_r H)\over 12H^{3\over 2}}\hat{\gamma}^r \hat{\gamma}_{\alpha}\left( 1 +\hat{\gamma}_0\hat{\gamma}_1\hat{\gamma}_2\right) \hat{\epsilon}, \nonumber \\
\delta \hat{\psi}_r &=& \partial_r \hat{\epsilon} - {(\partial_r H)\over 6H}\hat{\gamma}_0\hat{\gamma}_1\hat{\gamma}_2 \hat{\epsilon}.
\eea
The solution for this system of equations is again of the form,
\be
\epsilon=H^{-1/6}\epsilon_0.
\ee
where $\epsilon_0$ is a constant spinor  with only the expected $16$ real degrees of freedom.

To study the the supersymmetries in the near horizon limit for the eleventh dimensional case, we consider the relevant near horizon limit, giving as a result on the killing spinor equation
\bea
\partial_{\alpha} \hat{\epsilon} -\frac{u}{k^{1/2}}\hat{\gamma}_{\alpha}(1 -\hat{\gamma}_u)\hat{\epsilon} =0, \nonumber \\
\partial_u \hat{\epsilon} -\frac{1}{2u}\hat{\gamma}_u\hat{\epsilon}=0, 
\eea
which tell us that also in this case we get enhancement of supersymmetry.


\section{T-duality and Near Horizon Geometries} 


One of the questions that naturally arise, once we have a duality between QFT and string theory on given backgrounds, is the meaning of pure stringy symmetries from the point of view of the QFT. For example in the celebrated AdS/CFT correspondence the $SL(2,Z)$ S-duality of the Type IIB-string can be interpreted as the string version of the well known Monton Olive $SL(2,Z)$ duality of $N=4$ SYM \cite{sus1}. The quantum field theory meaning of T-duality is however a bit less clear since T-duality interpolates different D-branes and therefore tends to define maps between Yang Mills theories in different dimensions. Moreover T-duality transformation generally produce explicit breaking of supersymmetry.

The simplest possible way to address this questions on T-duality in a purely quantum field theoretical framework is of course by defining the quantum field theory using the near horizon limit of D-branes metrics and working out in this limit T-duality transformations.

To begin with, let us start considering a D3-brane living in a ten dimensional space time with one of the orthogonal coordinates compactified on a circle $S^1$ of radius $R$. The corresponding metric is given by
\bea
ds_3^2&=& H(r,R)^{-1/2}(dx_{||}^2)+H(r,R)^{1/2}(d(\theta R)^2+dr^2+r^2d\Omega_4) \nonumber \\
e^\phi&=&g,
\label{d3b}
\eea
with $x_{||}$ standing for world volume coordinates and the harmonic function $H$ given by
\be
H(r,R)=1+gl_s^4\sum_{n=1}^N\sum_{n_i}\left(\frac{Q_n}{[(y_9-y_{n_9}+n_9R)^2+\mid r-r_n \mid^2]^{4/2}}\right),
\label{har1}
\ee
where $g$ is the string coupling constant, $l_s$ is the string length, $Q_n$ is the charge of the D3-brane, $R$ is the radius of the circle $S^1$ and $r$ is the radius on spherical coordinates for the $\Re^5$ space time. From now on we will ignore any constant and will work with meaningful variables on the discussion, and always at large N.

At distances much longer than $R$, we can Poisson resume the expression (\ref{har1}) obtaining
\be
H(r,R)=1+{l_s^4g \over Rr^3}.
\label{har2}
\ee
This is a good solution as far as $R$ is small enough.

I we are interested in the near horizon limit \cite{mal1} of this metric we would be forced to defined this limit by performing a double ``blow up'', namely
\bea
\left ( \alpha'\rightarrow 0\;\;\;,\;\;\;U\equiv\frac{r}{\alpha'}= constant\;\;\;,\;\;\;v\equiv\frac{R}{\alpha'}=constant\right ).
\label{nh1}
\eea
Notice that the new variable $v$ correspond properly speaking to a blow up of the point $R=0$ in the moduli of the target space time metric (\ref{d3b}) with the harmonic function of (\ref{har2}). After performing the blow up (\ref{nh1}) we get the metric
\bea
ds_3^2&=& \alpha'\left[ \frac{U^{3/2}v^{1/2}}{g^{1/2}}(dx_{||}^2) + \frac{g^{1/2}}{U^{3/2}v^{1/2}}dU^2+ \frac{g^{1/2}U^{1/2}}{v^{1/2}}d\Omega_4 +\frac{g^{1/2}v^{3/2}}{U^{3/2}}d\theta^2 \right]\nonumber \\
e^\phi&=&g.
\label{m2} 
\eea
Note that the dilaton field for this solution is constant. The topology of the space time (\ref{m2}) is that o a fibration of a circle $S^1$ and a sphere $S^4$ on the space defined on the coordinates $(x^\alpha,U)$ with the corresponding radius
\bea
R_{S^1}&=&({gv^3 \over U^3})^{1/4}, \nonumber \\ 
R_{S^4}&=&({ gU \over v})^{1/4},
\label{rad1}
\eea
which depend on the moduli $v$. It is easy to see that this space time admits only $16$ real supercharges, the simples way to understand this is observing that $1 \over 2$ of the Killing spinors for the near horizon geometry of the D3-brane are projected out once we compactify a transverse direction (we will come back to this point later on). 

From the QFT point of view we should expect the metric (\ref{m2}) to be related to a SYM in $3+1$ with $16$ real supercharges and with a peculiar R-symmetry given by the isometries of $S^4 \otimes S^1$. The type of strings living on the space time (\ref{m2}) is Type IIB.

Consider next the candidate for a T-dual geometry, namely the D4-brane compactified on a $S^1$ of radius $R$, on its world volume. This solutions is given by
\bea
ds_4^2&=& H^{(-1/2)}(dx_{||}^2+d(\theta R)^2)+H^{(1/2)}(dr^2+r^2d\Omega_4) \nonumber \\
e^\phi&=&gH^{-1/4}, 
\eea
with $x_{||}$ expanding the non-compact four dimensional part of the world volume of the D4-brane, and the corresponding harmonic function
\be
H=1+\frac{gl_s^3}{r^3}.
\ee
Taking the near horizon limit, defined by again a double blow up
\bea
\left ( \alpha'\rightarrow 0\;\;\;,\;\;\;U\equiv\frac{r}{\alpha'}=constant\;\;\;,\;\;\;g_5^2\equiv g\alpha'^{1/2}=constant \right ),
\eea
we get the metric
\bea
ds_4^2&=& \alpha'\left[\frac{U^{3/2}}{g_5}(dx^2_{||}+(d\theta R)^2 )+\frac{g_5}{U^{3/2}}dU^2+g_5U^{1/2}d\Omega_4\right] \nonumber \\
e^\phi&=&g_5^{3/2}U^{3/4}.
\label{nh2}
\eea
Note that this time the dilaton is not constant. The number of supersymmetries in this brane is again $16$ real supercharges, however the QFT interpretation is a bit different, In this case we have a $D=5$ SYM theory living on $\Re^4\otimes S^1$. The corresponding radius of the  four-sphere and the circle are
\bea
R_{S^4}&=&(g_5^2U)^{1/4}, \nonumber \\ 
R_{S^1}&=&(\frac{U^{3}R^4}{g_5^2})^{1/4}.
\label{rad2}
\eea
\begin{figure}
\begin{center}
\includegraphics[angle=-90,scale=0.4]{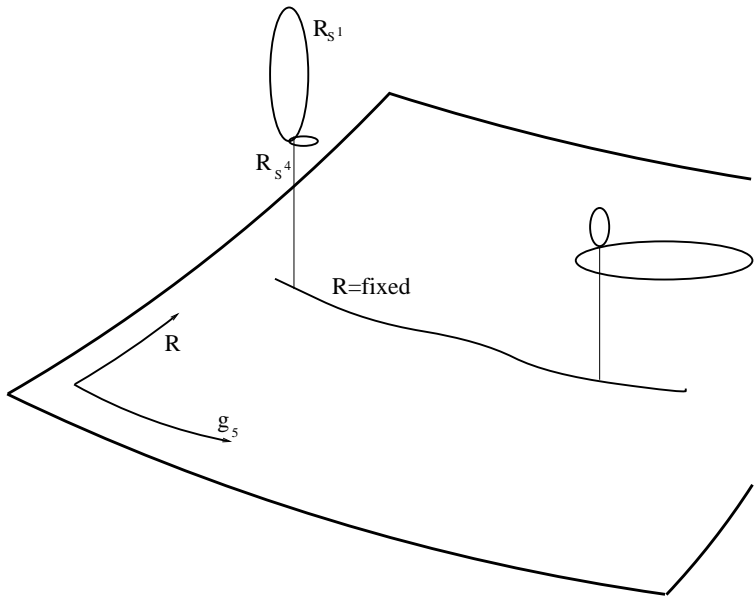}
\caption{Moduli space for near horizon D4-brane.}
\label{fig12}
\end{center}
\end{figure}
Before proceeding any further, let us clarify the picture we have (see fig. \ref{fig12}). In the D3-brane case we obtained the near horizon geometry as the result of a limit where it was left one of the moduli $g$ constant but we allowed $R$ to varied such that, at $R=0$ we create a divisor $v$. This two variables define our moduli $(v,g)$. The resulting geometry is that of a base space expanded by the coordinates $(x,U)$ and fibers $S^4$ and $S^1$ with the radius of equation (\ref{rad1}). In the D4-brane, we obtained the near horizon geometry as the result a another limit where one of the moduli $R$ is maintained constant while the other $g$ varies such that at infinite point we create a new divisor $g_5$. This two variables define the new moduli $(R,g_5)$. Again the geometry obtained is that of a base manifold expanded by the coordinates $(x,U)$ and fibers $S^4$ and $S^1$ with the radius of equation (\ref{rad2}). In order to identify both metrics (\ref{nh1},\ref{nh2}) by T-duality we most require the following relation between the different moduli,
\bea
g={g_5^2 \over R}, \nonumber \\
v={1 \over R}.
\label{td1}
\eea
Provided that this relation holds, we can perform the T-duality transformation following the normal Buscher rules \cite{bus1}. In principle we could run into difficulties if there appears to many singular point on the fibration, so that T-duality could loose its natural meaning. This T-dual map is well defined all over the base manifold, Actually we only have singularities at $U=0$ and $U=\infty$ but both points are related rather to wrong coordinate patches than real singularities. Therefore this T-dual map is a very trivial example of fiber-wise T-duality \cite{asp1}. The map described above is defined by T-duality, and effectively acts between the two moduli.

By now we have a neat relation between the bare coupling constants of both gauge theories in the two near horizon metrics. On the other hand it is usually associated to the dilaton behavior, the value of the corresponding running coupling constant i.e. the effective gauge coupling constant. Therefore to obtain the effective coupling of the compactified gauge theory on the world volume of the D4-brane, we considered the ratio of the effective coupling constant of the five dimensional gauge theory $g_5$ squared, with the effective radius of compactification namely the radius of the $S^1$ given on equation (\ref{rad2}), hence we get
\be
g_{4_{eff}}^2\equiv {g_{5_{eff}}^2 \over R_{eff}}.
\label{geff}
\ee
Then after solving for the moduli variables $(v,g)$ we obtain
\be
 g_{4_{eff}}^2=g.
\label{gf}
\ee
Therefore the effective coupling constant of the gauge theory we are studying from the point of view of the near horizon D4-brane has the same running as the the gauge theory on the near horizon D3-brane, as it should be expected invoking its duality relation. Note that the equation (\ref{geff}) was obtained by plausibly physical relations, however this equation is nothing more than the changing rule for the dilaton under T-duality!.

On the other hand the super Yang Mills theory on the D4 brane is not renormalizable, we can trust it only at low energies. This aspect of the gauge theory can be seen from the gravitational point of view. Note that for this geometry (\ref{nh2}) the dilaton grows for large $U$. Actually, we can trust on this solution as long as $U \geq 1/g_5^2$, after this point we should think in terms of M-theory.

The other possibility we have is to consider the D3-brane wrapped on a circle $S^1$ of radius $R$. This time the near horizon geometry is defined by the limit
\be
\left ( \alpha'\rightarrow 0\;\;\;,\;\;\;U\equiv\frac{r}{\alpha'}= constant \right ),
\ee
where ($g,R$) are kept constant on the process. Note that this time we don't have the double blow up of the above cases. The resulting metric is given by
\bea
ds_{3}^2&=& \alpha'\left[ \frac{U^2}{g^{1/2}}( d(R\theta)^2+dx_{||}^2)+\frac{g^{1/2}}{U^2}dU^2 +g^{1/2}d\Omega_{5})\right] \nonumber \\
e^\phi&=&g.
\label{nh3}
\eea
Again the dilaton is constant, and the topology is that of a fibration of a circle $S^1$ and the sphere $S^5$, on the space defined by the coordinates ($x^{\alpha},U$), with the corresponding radius
\bea
R_{S^1}&=&{UR \over g^{1/4}}, \nonumber \\ 
R_{S^5}&=&g^{1/4}.
\label{rad3}
\eea
This configuration also only admits 16 real supercharges. The situation on the field theory should be that of a $D=4$ SYM theory living on $\Re^3\otimes S^1$ with 16 real supercharges and R-symmetry contained on $SO(6)$.

The T-dual near horizon background is the result of first, a Poisson resume of the D2-brane solution with a transverse direction compactified on a small circle $S^1$ of radius $R$, second its  near horizon limit defined by the triple blow up, here showed
\bea
 \alpha'\rightarrow 0\;\;\;&,&\;\;\;U\equiv\frac{r}{\alpha'}=constant, \nonumber \\
v\equiv { R \over \alpha'}\;\;\;&,&\;\;\;g_3^2\equiv g\alpha'^{-1/2}=constant. 
\eea 
The resulting metric and dilaton are given by
\bea
ds_{2}^2&=& \alpha'\left[ \frac{U^2v^{1/2}}{g_3}(dx_{||}^2)+\frac{g_3}{U^2v^{1/2}}dU^2+{g_3 \over v^{1/2}}d\Omega_{5} + \frac{g_3v^{3/2}}{U^2}d\theta^2 \right] \nonumber \\
e^\phi&=& {g_3^{5/2} \over Uv^{1/4}}.
\label{nh4}
\eea
This time we have space time with 16 real supercharges, the metric defines a fibration on a circle $S^1$ and the sphere $S^5$ on the base space expanded by the coordinates ($x^\alpha,U$) and the field theory point of view should be that of a SYM theory on $2+1$ dimensions, with 16 real supercharges, and R-symmetry contained in $S0(6)\otimes U(1)$. The corresponding radius of the fibers are 
\bea
R_{S^1}&=&{g_3^{1/2}v^{3/4} \over U^{1/2}},\nonumber \\ 
R_{S^5}&=&{g_3^{1/2} \over v^{1/4}}.
\label{rad4}
\eea
To perform the T-duality map between this two metrics, we require to identify the moduli as follows,
\bea
g={g_3^2 \over v}, \nonumber \\
R={1 \over v}.
\label{td2}
\eea
Similar remarks about the validity of T-duality of the D3-brane and the D4-brane are applicable to this case.

In general, we can consider the above type of compactification of the D3-brane on a $T^{3-s}$, which take us to the near horizon D(s)-brane on a $T^{3-s}$ in the perpendicular coordinates. The metric of these geometries looks like $AdS_{s+2}\otimes S^5\otimes T^{3-s}$. When the holography map between the gauge theories and the bulk geometries is defined \footnote{Recall that the holographic conjecture of Maldacena only applies within a range of validity of $U$ for the case p different from 3}, the radius of the torus is very small. Also we found that when $R\rightarrow \infty$ we recover on these D(s)-branes, the full range of running for $U$ ($0,\infty$), while for small $R$ we are forced to stay at big values of the holographic variable $U$. 

It is well known that T-duality breaks supersymmetry in some cases \cite{alv1}, our previous metrics are not exceptions to this phenomenon. Note that we are relating theories with 16 real supercharges, but we alredy showed that the near horizon D3-brane shows enhancement of supersymmetry. The matching condition for the number of supersymmetries is given by the fact that the compact direction on the D3-brane (on both cases) eliminates the possibility of that enhancement, while the other Dp-branes don't show the enhancement at all.

It is important to notice that the supersymmetries which are broken by T-duality correspond to those which get enhanced in the particular case of the D3-brane, i.e. the bilinears associated with the broken Killing spinors, are the conformal Killing vectors.

In our previous analysis we have consider the two dual pairs (D3/D4) and (D2/D3). Both star with the D3-brane with the only difference that the compactified dimension is or not on the transversal direction. Let us study a bit more carefully both pairs. In the (D3/D4) case the D4-metric is characterized by two different moduli $(g_5,R)$. After the work in Matrix theory \cite{sei1}, it is natural to interpret $g_5$ as related to the eleventh dimension of M-theory and therefore to consider this metrics as coming from a compactification of M-theory on a $T^2$, with sizes determined by the two moduli $g_5$ and $R$. Also it is well known \cite{sus2} that in the limit where the volume of the two torus goes to zero, we should recover type IIB theory. This mechanism implies the dynamical generation of a ``quantum'' dimension with the corresponding Kaluza Klein modes associated with the menbrane wrapped on the two torus. When we apply this mechanism to our case it is natural to expect to get in the limit of zero volume for the two torus ($Vol(T^2)\rightarrow 0$) the type IIB D3-brane metric of equation (\ref{m2}). Using the relations (\ref{rad1}) and (\ref{td1}) we observe that the limit $Vol(T^2)\rightarrow 0$ corresponds to a ten dimensional type IIB theory, where the extra ``quantum'' dimension is the $S^1$ circle in the limit $v\rightarrow\infty$, with the radius given by (\ref{rad1}). The up lifted of the D4-brane to M-theory give us a M5-brane wrapped on a circle determined by the value of $g_5$. In addition to this, in our case we wrap the M5-brane on another circle defining the two torus characterized by the two moduli $(g_5,R)$. In the limit $Vol(T^2)\rightarrow 0$ what we get is the M5-brane wrapped on a two torus of zero volume, that produce a D3-brane with the extra dimension defining the transversional circle in the metric (\ref{m2}).

Hence the theory on the D4-brane gets embedded in the six dimensional ($2,0$) theory on the M5-brane. As it is well known for the M5-brane we get enhancement of supersymmetry and therefore we can say that the D4-brane theory will flow to a conformal point in strong coupling. {\it In other words what we observed is that once we break the superconformal generators by T-duality the resulting theory naturally flows to recover the supersymmetry by up lifting to M-theory}. 

Let us now consider the T-dual pair (D2/D3). This is very similar to the previous case. In the D2-brane metric the moduli is characterized by $(g_3,v)$, which again should by interpreted as M-theory compactified on a two torus of size $g_3$ and $v$. In the limit when the volume of the two torus goes to zero, we should recover the type IIB-picture by exactly the same mechanism described above. The D2-brane is now up lifted to a M2-brane but contrary to what happens in the (D3/D4) case, the extra ``quantum'' dimension becomes now part of the world volume dimension of the T-dual D3-brane. More precisely what we observed is that the compact ``world volume'' dimension in the metric (\ref{nh3}) is the extra ``quantum'' dimension in the type IIB that we get when we compactify M-theory on the two torus characterized by ($g_3,v$). {\it In other words what  we observed, is that the T-dual description in  (\ref{nh3}) of the uplifted D2-brane is a three dimensional theory becoming four dimensional for ``strong coupling'' $v\rightarrow 0$ in equation (\ref{td2})}.

As before with the (D3/D4) pair we also observe here that the theories flow to reach superconformal invariance. Notice that this two dual saturate the known examples of superconformal theories namely the D3-brane, M2-brane and M5-brane. The (D3/D4) pair is related to the M5-brane and the (D2/D3) pair to the M2-brane.
\begin{figure}
\begin{center}
\includegraphics[angle=-90,scale=0.4]{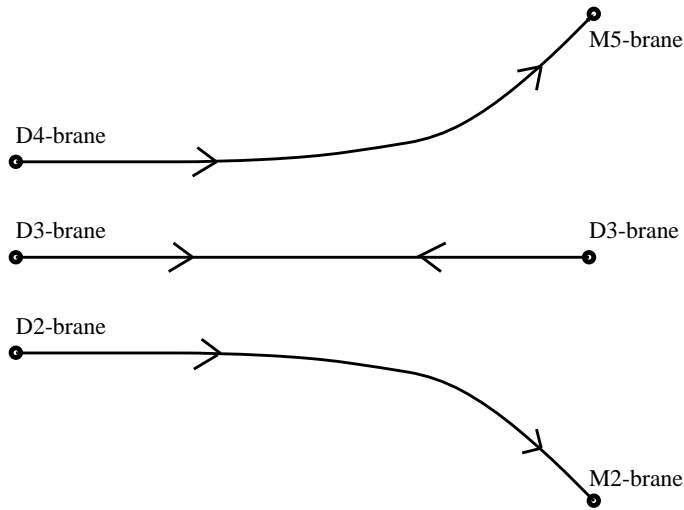}
\caption{Flow to the superconformal QFT.}
\label{fig13}
\end{center}
\end{figure}
The previous conjecture is in contrast to the mechanism suggested in \cite{wit1} for solving the cosmological constant problem. In that we can start with a three-dimensional theory that is expected to flow in strong coupling to a four-dimensional theory. Massive particles in three dimensions are associated with conical geometries, when some amount of supersymmetry is broken. The suggested solution to the cosmological constant problem, is based on the assumption that these supersymmetries are not restore in the strong coupling four-dimensional limit. In our case, we  have simply studied supersymmetry generators associated with conformal transformations that are the ones naturally broken by the action of T-duality, and we find they are restored in the up lifted ``M-theory'' limit.  

The general picture emerging from the previous discussion is that once we start with a superconformal theory, T-duality generally breaks the supersymmetries associated with the superconformal transformations, however the T-dual theory tends to flow to recover these supersymmetries broken by T-duality  up lifting to M-theory.

To be more precise, starting with a D3-brane with the world volume compactified on a circle of radius $R$, we break for finite radius the supersymmetries associated with those Killing spinors depending on world volume coordinates. Those are associated with the enhanced  supersymmetry. In order to decide if T-duality breaks or not supersymmetry, we perform a T-duality to a D2-brane. Once we have done that, we send the radius $R$ to infinite. In this limit we recover for the D3-brane the whole superconformal algebra, then if T-duality is not braking supersymmetry we should find that the T-dual of  the $R\rightarrow\infty$ limit possesses enhanced superconformal invariance. In fact this is what happened. By relation (\ref{td2}), when $R\rightarrow\infty$ then $v\rightarrow 0$ and $g_3^2\rightarrow 0$ for finite $g$, but $g_3^2$ can be interpreted as $1/\Delta$ for $\Delta$ the size of the eleventh dimension. Thus the D2-brane becomes uplifted to M2-brane recovering the superconformal transormations. In a certain sense M-theory is there to work out the breaking of supersymmetry induced by T-duality.

\vspace{12pt}
{\bf Acknowledgements}
\vspace{12pt}

We are grateful for stimulating discusions with E. Alvarez and T. Ortin. This work has been partially supported by European Union TMR program FMRX-CT96-0012 {\it Integravility, Non-pertubative  Effects, and Symmetry in Quantum Field Theory} and by the spanish grant AEN96-1655. The work of P.Silva was also suported by the Venezuelan goverment.


\begin{thebibliography}{99}

\bibitem{gsw}  M.B. Green, J.H. Schwarz and E. Witten, {\it Superstring Theory} 		(Cambridge U. Press 1987).
\bibitem{pol1} J. Polchinski {\it String theory} Vol. I and Vol. II 
	         (Cambridge U. Press 1998).
\bibitem{kis1} E. Kiritsis, {\it Introduction to Superstring Theory.} Leuven: 
		 Leuven University, to be publish,  hep-th/9709062.
\bibitem{jap1} A. Giveon, M. Porrati and E.Ravinivici, {\it Phys. Rep.} 
		 {\em 244},77 (1994).
\bibitem{pol2} J. Polchinski, {\it Phys. Rev. Lett.}{\em 75},4724 (1995).
\bibitem{fri1} D. Friedam, E. Martinec and S. Shenker, {\it Nucl. Phy.} 
		 {\em B271},93 (1986).
\bibitem{lei1} J. Polchinski, In {\it Fields, Strings and Duality. TASI 1996} 
		 eds. C. Efthimiou and B. Greene, 293.
\bibitem{mal2}  O. Aharony, S. S. Gubser, J. Maldacena, H. Ooguri, Y. Oz, 
		 hep-th/9905111.
\bibitem{kel1} K. S. Stelle, hep-th/9803116.
\bibitem{berg1} E. Bergshoeff, C. M. Hull and T. Ortin {\it Nucl.Phys.} 
		 {\em B451.} 547 (1995).
\bibitem{sus1} L. Susskind, hep-th/9611164
\bibitem{mal1} J. Maldacena, hep-th/9711200\\
	       N. Itzhaki et all, hep-th/9802042
\bibitem{bus1} T.H. Buscher,{\it Phys. Lett.} {\em B194} 59 (1987). \\
	       T.H. Buscher, {\it Phys. Lett.} {\em B201} 466 (1988) \\
               E. Alvarez, L. Alarez-Gaume and Y. Lozano, hep-th/9406206 
\bibitem{asp1} P.Aspinwall and R. Plesser, hep-th/9905036
\bibitem{alv1} E. Alvarez, L. Alvarez-Gaume and I. Bakas, hep-th/9507112\\
	       E. Alvarez, L. Alvarez-Gaume and I. Bakas,hep-th/9510028
\bibitem{ber1} E. Bergshoeff, hep-th/9611099
\bibitem{sei1} N. Seiberg, hep-th/9705117\\
	       M. Berkooz, M. Rozali an N. Seiberg, hep-th/9704089\\
	       M. Rozali, hep-th/9702136
\bibitem{sus2} S. Sethi and L. Susskind, hep-th/9702101
\bibitem{wit1} E. Witten, hep-th/9409111\\
	       E. Witten, hep-th/9506101
\end{thebibliography}
\end{document}